\documentclass[leqno,12pt]{article}
\usepackage{latexsym}
\renewcommand{\epsilon}{\varepsilon}

\usepackage{amsmath}
\usepackage{amssymb}
\usepackage{ntheorem}
\usepackage[ansinew]{inputenc}
\usepackage{graphicx}
\usepackage{epsfig}
\usepackage{caption}
\usepackage{url}
\usepackage[authoryear]{natbib}
\usepackage{amssymb}
\usepackage{setspace}
\usepackage{color}
\usepackage{amsmath}
\usepackage{graphicx}

\newcommand{\be}{\begin{eqnarray}}
\newcommand{\ee}{\end{eqnarray}}
\newcommand{\bea}{\begin{eqnarray*}}
\newcommand{\eea}{\end{eqnarray*}}

\begin{document}
\newtheorem{theorem}{Theorem}[section]
\newtheorem{corr}[theorem]{Corollary}
\newtheorem{lemma}[theorem]{Lemma}
\newtheorem{rem}[theorem]{Remark}
\newtheorem{definition}[theorem]{Definition}
\newtheorem{assump}{Assumption}
\newtheorem{exam}[theorem]{Example}
\title{Smooth backfitting in additive  inverse regression }

\author{Nicolai Bissantz, Holger Dette, Thimo Hildebrandt  \\
Ruhr-Universit\"at Bochum \\
Fakult\"at f\"ur Mathematik \\
44780 Bochum \\
Germany \\
{\small email: nicolai.bissantz@ruhr-uni-bochum.de }\\
{\small  \qquad  holger.dette@ruhr-uni-bochum.de}\\
{\small \quad thimo.hildebrandt@ruhr-uni-bochum.de} \\
{\small FAX: +49 234 32 14559}\\
}

 \maketitle

\begin{abstract}
We consider the problem of estimating an additive regression function in an inverse regression model with a convolution type operator. A smooth backfitting procedure is developed and asymptotic normality of the resulting estimator is established. Compared to other methods { for the estimation in additive models} the new approach  neither requires observations on a regular grid nor the estimation of the joint density of the predictor. It is also demonstrated by means of a simulation study that the backfitting estimator outperforms the marginal integration method at least by a factor two with respect to the integrated mean squared error criterion.
\end{abstract}

Keywords:  inverse regression; additive models; curse of dimensionality; smooth backfitting

Mathematical subject classification: Primary: 62G20; Secondary 15A29

\section{Introduction} \label{sec1}
\def\theequation{1.\arabic{equation}}
\setcounter{equation}{0}

In this paper we  consider the  regression model
\begin{eqnarray}
Y_{k} = g(\textbf{X}_{k})+\epsilon_{k} \label{model2}
\qquad k \in \{1,...,N\},
\end{eqnarray}
where $\epsilon_1,...,\epsilon_N$ are independent identically distributed random variables
 and $\textbf{X}_1,\dots, \textbf{X}_N$ are independent identically distributed $d$-dimensional predictors with components $\textbf{X}_k = (X_{k,1},\dots,X_{k,d})^T$ $(k=1,\dots,N)$. We assume that the function $g$ is related to a signal $\theta$ by a convolution type operator, that is
\begin{eqnarray}
g(\textbf{z})= \int_{\mathbb{R}^d} \psi(\textbf{z}-\textbf{t})\theta(\textbf{t}) d\textbf{t}, \label{conv}
\end{eqnarray}
where $\psi:\mathbb{R}^d\rightarrow \mathbb{R}$
 is a known function with $\int_{\mathbb{R}^d} \psi  (\textbf{t}) d \textbf{t}=1$. The interest of the experiment is the nonparametric estimation of the signal $\theta$.
 Models of the type  \eqref{model2} and \eqref{conv} belong to the class of inverse regression models and have important
applications  in the recovery of images from  astronomical telescopes or fluorescence microscopes in biology.
Deterministic inverse regression models have been considered for a long time in the literature   [\cite{enghanneu1996}, \cite{saitoh1997}].
However, in  the last decade
statistical inference  in ill-posed problems has become a very active field of research
[see \cite{berbocdes2009},  \cite{kaisom2005}  for a Bayesian approach and \cite{mairuy1996}, \cite{cavalier2008} and \cite{bishohmunruy2007}
for nonparametric methods].

 While most of these methods have been developed for models with a  one-dimensional predictor, nonparametric estimation in the multivariate setting is
  of practical importance because in many applications one has to deal with an
at least two-dimensional predictor. A typical example is image reconstruction since a picture
is a two-dimensional object. Also in addition to the spatial dimensions, the
   data might depend on the time thus introducing a third component.   For a multivariate predictor
   the estimation of the signal $\theta$ in the inverse regression model \eqref{model2} is
  a much harder
  problem due to the curse of dimensionality. In direct regression usually qualitative assumptions regarding the signal such as  additivity or multiplicativity are made, which allow the
 estimation of the regression function at reasonable rates  [see \cite{linnie1995}, \cite{mamlinnie1999}, \cite{carhaemam2002},
\cite{henspe2005}, \cite{niespe2005}]. In the present paper we investigate the problem of estimating the signal $\theta$ in the inverse regression model with a convolution type operator under the additional assumption of additivity, that is
\begin{eqnarray} \label{VAM}
\theta (\textbf{x}) &=& \theta_0+\theta_1(x_1)+...+\theta_d(x_d),
\end{eqnarray}
 where ${\bf x}=(x_1,\dots,x_d)^T$.
 In a recent paper \cite{hilbisdet13}
  proposed an estimator of the signal $\theta$ if observations are available on a regular grid in $\mathbb{R}^d$. They also considered the case of a random design and investigated the statistical properties of a marginal integration type estimate with known density of the predictor. The asymptotic analysis of both estimates is based on these rather restrictive assumptions regarding the predictor $\textbf{X}$. A regular grid or explicit knowledge of the density of the predictor $\textbf{X}$ might not be available in all applications.  Moreover, estimation of this density in the marginal integration method cannot be performed at one-dimensional rates
    [see  \cite{hilbisdet13}].  In particular it changes the asymptotic properties of additive estimates such that the signal cannot be reconstructed with one-dimensional nonparametric rates. In the present paper we consider the construction of an estimate in the inverse additive regression model \eqref{VAM} with random design, which is applicable under less restrictive assumptions in particular without knowledge of the density of the predictor. For this purpose we combine in Section \ref{sec2} smooth backfitting [see \cite{mamlinnie1999}] with Fourier estimation methods in inverse regression models [{see \cite{digglehall93} or \cite{mairuy1996}]. Besides several advantages of the smooth backfitting approach observed in the literature in direct regression models [see \cite{niespe2005}], the backfitting methodology only requires  the estimation of the marginal densities of the predictor. As a consequence, the  resulting estimate does not suffer from the curse of dimensionality.
Section \ref{sec3} is devoted to the investigation of the asymptotic  properties of the new estimator, while we study the finite sample properties  by means of a simulation study in Section \ref{sec4}. In particular we demonstrate that the smooth backfitting approach results in estimates with an at least two times smaller integrated mean squared error than the marginal integration method. Finally, all proofs and technical arguments are presented in Section \ref{sec5}.

\section{Smooth backfitting in inverse regression} \label{sec2}
\def\theequation{2.\arabic{equation}}
\setcounter{equation}{0}

 Note that the linearity of the convolution operator  and assumption \eqref{VAM} imply that the function $g$ is also additive, and consequently the model \eqref{model2} can be rewritten as
\begin{eqnarray}\label{direct}
Y_k &=& g_0 + g_1(X_{k,1})+...+g_d(X_{k,d})+\epsilon_k,
\end{eqnarray}
 where $\textbf{X}_k = (X_{k,1},\dots,X_{k,d})^T$ and the functions $g_0, g_1,\dots,g_d$ in model \eqref{direct} are related to the components $\theta_0, \theta_1,\dots,\theta_d$ of the signal $\theta$ in model \eqref{VAM} by $g_0 = \theta_0$,
\be \label{gj}
g_j(z_j)= \int_{\mathbb{R}} \psi_j (z_j-t) \theta_j(t)dt \qquad j=1,\dots,d.
\ee
Here $\psi_j$ is the marginal of the convolution function $\psi$, that is
\be \label{marpsi}
\psi_j(t_j) = \int_{\mathbb{R}^{d-1}} \psi({\bf t}) d {\bf t_{-j}}
\ee
and $\textbf{t}=(t_1,...,t_d)^T \in \mathbb{R}^d,{\bf t_{-j}}=(t_1,\dots,t_{j-1}, t_{j+1},\dots,t_d)^T \in \mathbb{R}^{d-1}$. The estimation of the additive signal is now performed in several steps and combines Fourier transform estimation methods for inverse regression models [see \cite{digglehall93} or \cite{mairuy1996}] with the smooth backfitting technique developed for direct nonparametric regression models [see \cite{mamlinnie1999}].
\begin{itemize}
\item [(1)] We assume for a moment that the design density is known and denote  by $f_j$ and $F_j$ the density and cumulative distribution function of the $j$th marginal distribution of the random variable $\textbf{X}$. In a first step all explanatory variables are transformed to the unit cube by using the probability transformation in each component, that is
    \be \label{probtrafo}
    Z_{k,j}=  F_j(X_{k,j}) \qquad \qquad j=1,\dots,d; \quad \ k=1,\dots,N.
    \ee
    This transformation is necessary because of two reasons. On the one hand,  the asymptotic analysis of methods based on Fourier estimation requires with positive probability observations at points $\textbf{X}_k$ with a norm $\parallel \textbf{X}_k \parallel$ converging to infinity, because one has to estimate the Fourier transform of the function $g_j$ on the real axis. On the other hand,  the asymptotic analysis of the smooth backfitting method requires a distribution of the explanatory variables with a compact support. \\
    In practice the unknown marginal distributions of the predictor are estimated by standard methods and this estimation does not change the asymptotic properties of the statistic. We  refer to Remark \ref{rem1} for more details.
    \item [(2)] The transformation in Step (1) yields the representation
    \be \label{trans}
    Y_k = g_0+ g^*_1 (Z_{k,1}) + \dots + g^*_d (Z_{k,d}) + \varepsilon_k \, ; \qquad \qquad k=1,\dots,N,
    \ee
    where the functions $g_j^*$ are defined by $g^*_j = g_j \circ F^{-1}_j \ (j=1,\dots,d)$. We now use the smooth backfitting algorithm [see \cite{mamlinnie1999}] to estimate each function  $g^*_j$ in \eqref{trans} from the data $(Z_{1,1},...,Z_{1,d}, Y_1),\dots,(Z_{N,1},...,Z_{N,d},Y_N)$. This algorithm determines estimates of the components $g_0,g^*_1,\dots,g^*_d$ recursively, where $\hat g_0 = \overline Y. = \frac 1 N \sum_{k=1}^N Y_k$. For starting values $\hat g^{*(0)}_1,\dots,\hat g^{*(0)}_d$ we calculate for $r=1,2,\dots$ the estimators $\hat g^{*(r)}_1, \dots, \hat g^{*(r)}_d$ by the recursive relation
\begin{eqnarray} \label{backalg}
\hat g_j^{*(r)} (z_j) &=& \hat{g}^*_j(z_j)-\sum_{k<j} \int \hat g_k^{*(r)}(z_k)\Big[ \frac{\hat{p}_{jk}(z_j,z_k)}{\hat{p}_j(z_j)}-\hat{p}_{k,[j+]}(z_k)\Big]dz_k
\\ && -\sum_{k>j} \int g_k^{*(r-1)}(z_k)\Big[ \frac{\hat{p}_{jk}(z_j,z_k)}{\hat{p}_j(z_j)}-\hat{p}_{k,[j+]}(z_k)\Big]dz_k-g^*_{0,j} ~. \nonumber
\end{eqnarray}
Here
\begin{equation} \label{NW}
\hat{g}_j^* (z_j) = \frac {\sum_{k=1}^N L \Big( \frac {Z_{k,j} - z_j} {h_B} \Big) Y_k} {\sum_{k=1}^N L \Big( \frac {Z_{k,j} - z_j} {h_B} \Big)}
\end{equation}
denotes the one-dimensional Nadaraya-Watson estimator of the $j$th component (with kernel $L$ and bandwidth $h_B$),
 $\hat{p}_{jk}$  and $\hat{p}_j$ are the ($j,k$)th and $j$th marginals of the common kernel density estimator $\hat{p}$ for the density $p$ of the predictor $(Z_1,...,Z_d)^T$,   and
 we use the notation
\begin{eqnarray}\label{g0j} \nonumber
\hat{p}_{k,[j+]}(z_k) &=& \int \hat{p}_{jk}(z_j,z_k) dz_j \Big[\int \hat{p}_j(z_j)dz_j\Big]^{-1}
\\ g^*_{0,j} &=& \frac{\int \hat{g}_j^*(z_j)\hat{p}_j(z_j) dz_j}{\int \hat{p}_j(z_j)dz_j}.
\end{eqnarray}
  \item [(3)] Estimators of the functions $g_j$ in \eqref{direct} are now easily obtained by the transformation
  \begin{equation} \label{gsch}
  \hat g_j = \hat g^{*(r_0)}_j \circ F_j,
  \end{equation}
  where $ \hat g_j^{*(r_0)}$ denotes the estimator obtained after terminating the recursive relation \eqref{backalg} at step $r_0 \ (j=1,\dots,d)$.  In order to recover the signal $\theta_j$ from $\hat g_j$ we now  introduce the random variables
\begin{eqnarray}
 U_{k,j} = Y_k -\sum_{\substack{i=1 \\ i\not=j}}^d \hat{g}_i(X_{k,i}) - \hat g_0 \label{ukj}
\end{eqnarray}
and use the data  $(X_{1,j}, U_{1,j}),\dots, (X_{N,j}, U_{N,j})$ to estimate the $j$th component $\theta_j$ of the signal $\theta$ by Fourier transform estimation methods [see \cite{digglehall93} for example]. For this purpose we note that the relation \eqref{gj} implies for the Fourier transforms $\Phi_{g_j}$ and $\Phi_{\theta_j}$ of the functions $g_j$ and $\theta_j$ the relation
        $$
        \Phi_{\theta_j} = \frac {\Phi_{g_j}}{\Phi_{\psi_j}},
        $$
        where
        $$
        \Phi_{\psi_j} (w) = \int_\mathbb{R} \psi_j(x) e^{iwx}
dx        $$
         is the Fourier transform of the $j$th marginal of the convolution function.
       Now the Fourier transform $ \Phi_{g_j} (w) $  of the function $g_j$ is estimated by its empirical counterpart
        \be \label{gjest}
        \hat{\Phi}_{g_j}(w) = \frac {1}{N} \sum^N_{k=1} e^{iw X_{k,j}}  \frac {U_{k,j}}{\max \{  f_j(X_{k,j}), f_j (\frac {1}{a_N}) \}},
        \ee
        where $f_j$ is the density of the $j$th marginal distribution and $a_N$ is a real valued sequence converging to $0$ as $N \to \infty$. The estimator of $\hat \theta_j$ is now obtained from a ``smoothed''  inversion of the  Fourier transform, that is
        \be \label{tjest}
        \hat \theta_j(x_j) = \frac {1}{2 \pi} \int_{\mathbb{R}} e^{-iwx_j} \Phi_K (hw) \frac {\hat \Phi_{g_j}(w)}{\Phi_{\psi_j} (w)} dw,
        \ee
        where $\Phi_K$ is the Fourier transform of a kernel $K$ and $h$ is a bandwidth converging to $0$ with increasing sample size.
        \item [(4)] Finally, the additive estimate of the signal $\theta$ is given by
\begin{eqnarray}
\hat{\theta} (\textbf{x}) =\hat{\theta}_0+ \hat{\theta}_1(x_1)+...+\hat{\theta}_d (x_d) , \label{thetaaddbfrd2}
\end{eqnarray}
where $\hat \theta_0 = \hat g_0 = \overline{Y}.$ and $\hat \theta_j$ is defined in \eqref{tjest} for $j=1,\dots,d$.
\end{itemize}

\begin{rem} \label{rem1}  \quad
\begin{itemize}
\item [ {\rm (a)}] {\rm Note that we use the term $\max \{ f_j (X_{k,j}), f_j (\frac {1}{a_N}) \}$ in the denominator of the estimate \eqref{gjest} instead of the more intuitive term $f_j (X_{k,j})$. This ``truncation'' avoids situations where the denominator becomes too small, which would yield unstable estimates with a too large variance.}
\item [ {\rm (b)}]{\rm In practical applications knowledge of the marginal distributions might not be available and in this case the transformation \eqref{probtrafo} can be achieved by
    \be \label{pract}
    \hat Z_{k,j} = \hat {\mathbb{F}}_j (X_{k,j}); \qquad \qquad j=1,\dots,d; \quad k=1,\dots,N,
    \ee
    where for $j=1,\dots,d$
    $$
    \hat {\mathbb{F}}_j(x) = \frac {1}{N+1} \sum^N_{k=1}\mathbb{I}  \{ X_{k,j} \leq x \}
    $$
    denotes the empirical distribution function of the $j$th components $X_{1,j},...,X_{N,j}$. Similarly, the density $f_j$ in \eqref{gjest} can be estimated by kernel density methods,
    that is
    \be \label{dens}
    \hat f_j (x_j) = \frac {1}{Nh_{d,j}} \sum^N_{k=1} M \Bigl ( \frac {X_{k,j}- x_j}{h_{d,j}} \Bigr ) \: ; \qquad j=1,\dots,d,
    \ee
    where $M$ denotes a kernel and $h_{d,j}$ is a bandwidth proportional to $N^{-1/5}$. We note that the estimators $\hat{\mathbb{F}}_j$ and $\hat f_j$ converge uniformly to $F_j$ and $f_j$ at rates $(\frac {\log \log N}{N})^{1/2}$ and $(\frac { \log N}{Nh_{d,j}})^{1/2}$, respectively [see \cite{vaart1998}, \cite{gingui2002}]. The rates of convergence in inverse deconvolution problems are slower and consequently the asymptotic properties of the estimates $\hat \theta_j$ do not change if $f_j$ and $F_j$ are replaced by their empirical counterparts $\hat f_j$ and $\hat{\mathbb{F}}_j$ defined in \eqref{pract} and \eqref{dens}, respectively.
    }
    \end{itemize}
    \end{rem}

\section{Asymptotic properties} \label{sec3}
\def\theequation{3.\arabic{equation}}
\setcounter{equation}{0}

In this section we investigate the asymptotic properties   of the estimators defined in Section \ref{sec2}. In particular we establish weak convergence. For this purpose we require the following assumptions
\begin{itemize}
\item [(A1)] The kernel $L$ in the Nadaraya-Watson estimator $\hat g^*_j$ in the backfitting recursion \eqref{backalg} is symmetric, Lipschitz continuous and has compact support, say $[-1,1]$. The bandwidth $h_B$ of this estimator is proportional to $N^{- 1/5}$.
    \item [(A2)] $\mathbb{E}[|Y_j|^\alpha] < \infty$ for some $\alpha > \frac {5}{2}$.
    \item [(A3)] The functions $g_1,\dots,g_d$ in model \eqref{direct} are bounded and twice differentiable with Lipschitz continuous second order derivatives.
\item[{(A4)}] {\rm The Fourier transforms $\Phi_{\psi_{j}}$ of the marginals $\psi_{j}$ of the convolution function
$\psi$  satisfy
\begin{eqnarray*}
&&\int_{\mathbb{R}} \frac{|\Phi_K({w})|}{|\Phi_{\psi_{j}}(\frac{{w}}{h})|} d {w} \le C_1h^{-\beta_j}, \qquad
 \int_{\mathbb{R}} \frac{|\Phi_K({w})|^2}{|\Phi_{\psi_{j}}(\frac{{w}}{h})|^2} dw \sim C_2h^{-2\beta_j}, \\
 && \left\vert \frac{1}{h}\int \int e^{-iw(x-x_j)/h} \frac{\Phi_K(w)}{\Phi_{\psi_j}(\frac{w}{h})}dw \frac{f_j(x)}{\max\{f_j(x),f_j(\frac{1}{a_N}\}}dx \right\vert = o(h^{-2\beta-1})
\end{eqnarray*}
uniformly with respect to $x_j$
for some constants $\beta_j >0$ $(j=1,\dots,d)$ and constants  $C_1,C_2,C_3 >0$, where the constant $C_3$ does not depend on $x_j$.}
\item[{ (A5)}] The Fourier transform $\Phi_K$ of the kernel $K$ is symmetric and supported on the interval $[-1,1]$. Additionally there exists a constant
 $b \in (0,1]$ such that $\Phi_K({w})=1$ for all ${w} \in [-b,b],b> 0,$ and $|\Phi_K({w})| \le 1$ for all ${w} \in \mathbb{R}$
\item[{ (A6)}] {\rm The Fourier transforms $\Phi_{\theta_{1}},\ldots ,\Phi_{\theta_{d}}$ of the functions $\theta_{1},\ldots ,\theta_{d}$ in the additive model \eqref{VAM}  satisfy
\begin{eqnarray*}
\int_{\mathbb{R}} |\Phi_{\theta_{j}}({w})|  |{w} |^{s-1} d {w} < \infty \quad \mbox{for some} \quad s>1 \mbox{ and } j=1,...,d.
\end{eqnarray*}}
\item[{(A7)}] {\rm The functions $g_{1},...,g_{d}$ defined in model \eqref{gj} satisfy
\begin{eqnarray*}
\int_{\mathbb{R}} |g_{j}({z})| |  {z} | ^r d {z} < \infty \quad \mbox{for} \quad j=1,...,d
\end{eqnarray*}
for some $r>0$ such that  $a_N^{r-1} = o(h^{\beta_j+s})$.}
\item[{(A8)}]
 {\rm For each $N\in \mathbb{N}$ let $\textbf{X}_1,...,\textbf{X}_N$ denote independent identically distributed $d$-dimensional  random variables with marginal densities $f_1,...,f_d$ (which may depend on $N$) such that $f_j(x) \not= 0$ for all $x\in [-\frac{1}{a_N},\frac{1}{a_N}]$. We also assume that $F_j^{-1}$ exists, where $F_j$ is the distribution function of $X_{1,j}$. Furthermore we assume, that
 for sufficiently large $N \in \mathbb{N}$
\[f_j(x) \ge f_j(\frac{1}{a_N}) \quad \mbox{ whenever } \quad x \in [-\frac{1}{a_N},\frac{1}{a_N}],\]
  }
  for all $j=1,\dots,d$.
\item[{(A9)}] If $f_{ijk} (t_i , t_j | t_k)$ and $f_{ij} (t_i | t_j)$ denote the densities of the conditional distribution $\mathbb{P}^{X_i , X_j | X_k}$ and $\mathbb{P}^{X_i | X_j}$, respectively, we assume that there exist integrable functions (with respect to the Lebesgue measure), say $U_{ijk} : \mathbb{R}^2 \to \mathbb{R} ~,~ \eta_{ij} : \mathbb{R} \to \mathbb{R}$, such that the inequalities
\[ f_{ijk} (t_i , t_j | t_k) \le U_{ijk} (t_i , t_j) ~;~ f_{ij} (t_i | t_j) \le \eta_{ij} (t_i) \]
are satisfied for all $t_i , t_j , t_k \in \mathbb{R}$.
\end{itemize}

\begin{rem}
{\rm Assumption (A1) - (A3) are required for the asymptotic analysis of the backfitting estimator, while (A4) - (A8) are used to analyze the Fourier estimation methods used in the second step of the procedure.  In order to demonstrate that these assumptions are satisfied in several cases of practical importance we consider exemplarily Assumption (A4) and (A6).
\begin{itemize}
\item [(a)] To illustrate Assumption (A4) the convolution function $\psi$ and the kernel $K$ are  chosen as
\begin{eqnarray*}
\psi_j(x) &=& \frac{\lambda}{2}e^{-\lambda|x|}; \qquad  K(x) = \frac{\sin(x)}{\pi x},
\end{eqnarray*}
respectively. Furthermore  we choose $f_j$ as density of a uniform distribution on the interval $[-\frac{1}{a_N},\frac{1}{a_N}]$ and consider exemplarily the point $x_j=0$. Note that $\Phi_K(w) =  \mathbb{I}_{[-1,1]}(w)$. The integrals in (A4) are obtained by straightforward calculation, that is
 \begin{eqnarray*}
\int_{\mathbb{R}} \frac{|\Phi_K(w)|}{|\Phi_{\psi_j}(\frac{w}{h})|} dw &=& \int_{[-1,1]} \left(1+\frac{w^2}{h^2}\right) dw = \frac{2}{3h^2}+2
\end{eqnarray*}
 \begin{eqnarray*}
 \int_{\mathbb{R}} \frac{|\Phi_K(w)|^2}{|\Phi_{\psi_j}(\frac{w}{h})|^2} dw&=& \int_{[-1,1]} \left(1+\frac{w^2}{h^2}\right)^2 dw= \frac{2}{5h^4}+\frac{4}{3h^2}+2
\end{eqnarray*}
\begin{eqnarray*}
&& \frac{1}{h}\int_{[-1/a_N,1/a_N]} \int_{[-1,1]} e^{-iw(x-x_j^*)/h} \frac{\Phi_K(w)}{\Phi_{\psi_j}(\frac{w}{h})}dw \frac{f_j(x)}{\max\{f_j(x),f_j(\frac{1}{a_N}\}} dx
\\ &=& \frac{2}{ h}\int_{[-1/a_N,1/a_N]}  \frac{((h^2(x^2-2)+x^2)\sin(\frac{x}{h})+2hx\cos(\frac{x}{h}))}{hx^3} dx
\\ &=& \frac{-2a_N \cos(\frac{1}{a_Nh})+2a_N^2h\sin(\frac{1}{a_Nh})+2hSi(\frac{1}{a_Nh})}{h}
\end{eqnarray*}
and $Si(x)$ denotes the sine-integral $\int_0^x \frac{\sin(y)}{y} dy$.
This shows that condition (A4) is satisfied.
\item[(b)]In order to illustrate Assumption (A6) let $W^m(\mathbb{R})$ denote the Sobolev space of order $m \in \mathbb{N}$, then the assumption $\theta_j \in W^{s}(\mathbb{R})$ with $s \in \mathbb{N}\backslash\{1\}$ implies condition (A6). Conversely, if (A6) holds with $s \in \mathbb{N}\backslash\{1\}$, then $\theta_j$ is $(s-1)$ times continuously differentiable [see \cite{folland1984}].   In other words, (A6) is an assumption regarding the smoothness of the components of the signal $\theta_j \ (j=1,\dots,d)$.
\end{itemize}}
\end{rem}

Our main result, which is proved in the Appendix, establishes the weak convergence of the estimator $\hat \theta_j$  for the $j$th component of the additive signal in model \eqref{VAM}. Throughout this paper the symbol $\Rightarrow$ denotes weak convergence.

\medskip

\begin{theorem}
\label{theo1}
Consider the additive inverse regression model defined by \eqref{model2} - \eqref{VAM}.
If Assumptions (A1) - (A8)   are satisfied and additionally the conditions
\begin{eqnarray}
   N^{1/2}h^{\beta_j+1/2}f_j(\frac{1}{{a}_{N}})^{1/2} \rightarrow \infty \label{b1} \\
    N^{1/2}h^{3/2}f_j(\frac{1}{{a}_{N}})^{3} \rightarrow \infty, \quad N^{1/5}h^{s+\beta_j}f_j(\frac{1}{{a}_{N}}) \rightarrow \infty \label{b2}
\end{eqnarray}
are fulfilled, then a standardized version of the estimator $\hat{\theta}_j$ defined in
\eqref{tjest} converges weakly, that is
\begin{eqnarray*}
V_{N,j}^{-1/2} \left( \hat{\theta}_j(x_j)-\mathbb{E}[\hat{\theta}_j(x_j)]\right) \Rightarrow \mathcal{N}(0,1),
\end{eqnarray*}
where
\[ \mathbb{E}[\hat{\theta}_j(x_j)] = \theta_j(x_j) + o(h^{s-1}), \]
and the normalizing sequence is given by
\begin{eqnarray} \label{VNj}
V_{N,j} &=&\frac{1}{Nh^{2}(2\pi)^{2}}\int_{\mathbb{R}}  \left|  \int_{\mathbb{R}} e^{-iw (x_j-y)/h} \frac{\Phi_K(w)}{\Phi_{\psi_j}(\frac{w}{h})} dw\right|^2 \frac{(g_j^2(y)+\sigma^2)f_j(y)}{\max\{ f_j(y),f_j(\frac{1}{a_N})\}^2} dy
\end{eqnarray}
and satisfies
\begin{eqnarray} \label{VNjung}
N^{1/2}h^{\beta_j+1/2}f_j\left(\frac{1}{a_N}\right)^{1/2} \le V_{N,j}^{-1/2} \le N^{1/2}h^{\beta_j+1/2}.
\end{eqnarray}
\end{theorem}

As a consequence of Theorem \ref{theo1} we obtain the weak convergence of the additive estimate $\hat \theta$ of the signal $\theta$.

\begin{rem}\label{remarkoracle}
{\rm If all components except one would be known, it follows from Theorem 3.1 in \cite{hilbisdet13} that this component can be estimated at a rate $R_N$ satisfying
$$
\frac {c_1}{N^{1/2}h^{1/2+\beta_j}} \leq R_n \leq \frac{c_2}{N^{1/2} h^{1/2+ \beta_j}f_j(a^{-1}_N)}
$$
(with appropriate constants $c_1$ and $c_2$). Consequently, it follows from Theorem \ref{theo1} that the smooth backfitting operator $\hat \theta_j$ defined in \eqref{tjest} has an oracle property and estimates the $j$th component at the one-dimensional rate. }
\end{rem}

\begin{corr}
\label{corr} Consider the inverse regression model defined by \eqref{model2} - \eqref{VAM} and assume that the assumptions of Theorem \ref{theo1} are satisfied for all $j=1, \dots,d$.
Then  a standardized version of the the additive estimator $\hat{\theta}$ defined in
\eqref{thetaaddbfrd2}  converges weakly, that is
\begin{eqnarray*}
V_{N}^{-1/2} \left( \hat{\theta} (x)-\mathbb{E}[\hat{\theta} (x)]\right) \Rightarrow \mathcal{N}(0,1).
\end{eqnarray*}
Here
\[ \mathbb{E}[\hat{\theta} (x)] =\theta (x)+ o(h^{s-1}), \]
and the normalizing factor is given by $V_N= \sum_{j=1}^d V_{N,j}+ \sum_{1\le k \not = l \le d} V_{N,k,l}$, where $V_{N,j}$ is defined in \eqref{VNj},
\begin{eqnarray*}
V_{N,k,l} &=& \frac{1}{Nh^2(2\pi)^2}\int_{\mathbb{R}}   \int_{\mathbb{R}} e^{-iw (x_k -y)/h} \frac{\Phi_K(w)}{\Phi_{\psi_k}(\frac{w}{h})} dw \overline{\int_{\mathbb{R}} e^{-iw (x_l -z)/h} \frac{\Phi_K(w)}{\Phi_{\psi_l}(\frac{w}{h})} dw }
\\ &&\times \frac{(\sigma^2+g_k(y)g_l(z))f_{k,l}(y,z)}{\max\{ f_k(y),f_k(\frac{1}{a_N})\}\max\{ f_l(y),f_l(\frac{1}{a_N})\}} d(y,z),
\end{eqnarray*}
and $f_{k,l}$ denotes the joint density of the pair $(X_{k,1},X_{l,1})$. Moreover $V_N$ satisfies
\begin{eqnarray*}
N^{1/2}h^{\beta_{j^*}+1/2}f_{j^*}\left(\frac{1}{a_N}\right)^{1/2} \le V_{N}^{-1/2} \le N^{1/2}h^{\beta_{j^*}+1/2}.
\end{eqnarray*}
where $j^* = \mbox{argmin}_j h^{\beta_j} f_j (1/a_N)$.

\end{corr}

\section{Finite sample properties} \label{sec4}
\def\theequation{4.\arabic{equation}}
\setcounter{equation}{0}
In this section we briefly investigate the finite sample properties of the new backfitting estimators by means of a small
simulation study. We also compare the two estimators obtained by the marginal integration method with the backfitting estimator proposed in this paper.
All results are based on $500$ simulation runs. For the sake of brevity we concentrate on three models with a two-dimensional predictor
and two distributions for the predictor. To be precise we consider the models
\begin{eqnarray}
\theta(x_1,x_2) &=& \theta_1(x_1)+ \theta_2(x_2) = e^{-(x_1-0.4)^2}+e^{-(x_2-0.1)^2} \label{sig1},\\
\theta(x_1,x_2) &=& \theta_1(x_1)+ \theta_2(x_2) = x_1e^{-|x_1|}+ ( {1+x_2^2} )^{-1} \label{sig2} ,\\
\theta(x_1,x_2) &=& \theta_1(x_1)+ \theta_2(x_2) = e^{-|x_1|}+ ({1+x_2^2} )^{-1}, \label{sig3}
\end{eqnarray}
and assume that the convolution function is given by
\begin{eqnarray}
\label{conv2}
\psi ({x}_1,x_2 ) = \frac {9} {4} e^{-3(\vert x_1 \vert + \vert x_2 \vert)}.
\end{eqnarray}
Note that the signals in \eqref{sig1}  and \eqref{sig2} satisfy the assumptions posed in Section \ref{sec3}, while this is not the case for
the first component of the signal \eqref{sig3}.
For the  distribution of the explanatory variable  we consider  an  independent and correlated case, that is
\begin{eqnarray}
&& \mbox{a uniform distribution on the square } [1/a_N,1/a_N]^2
\label{des1}\\
&&  \mbox{a two-dimensional normal  distribution with mean }  {\bf 0}   \mbox{  and variance }
\label{des2}
~  \Sigma =\left(\begin{array}{rr} 1 & \frac{1}{\sqrt{2}} \\
       \frac{1}{\sqrt{2}}  & 1 \end{array}\right) 
\end{eqnarray}
The sample size is $N=701$,  the variance is given by $\sigma^2=0.25$ and for the sequence $a_N$ we used $0.5$.
  In the simulation the bandwidths are chosen in several (nested) steps. At first    the bandwidths $h_{d,j}$
 in \eqref{dens}    are calculated minimizing the mean integrated squared error of the density estimate. These bandwidths are used
 in the calculation of the mean integrated squared error of   the estimate $\hat{g}_j$ in \eqref{gsch}, which is then
 minimized with respect to the choice of $h_B$. The final step consists of a calculation of  the bandwidth $h$ minimizing
 the mean integrated squared error of the resulting inverse Fourier transform  \eqref{tjest}. In practice
 this procedure of the mean squared error requires knowledge of the quantities  $f_j$, $g_j$ and
for  a concrete application we recommend to mimic these calculations by cross validation.

 In Figures \ref{fig1} -
\ref{fig3}
we present the estimated mean curves for both components corresponding to model \eqref{sig1}  - \eqref{sig3} respectively. Upper parts  of the tables show the results for 
 independent components of the predictor, where the case of correlated explanatory variables is displayed in the lower panels. The figures also contain the (pointwise)
estimated $5\%$ and $95\%$-quantile curves to illustrate the variation of the estimators.
We observe that in  models  \eqref{sig1} and \eqref{sig2} both  components are estimated with reasonable
precision [see Figure \ref{fig1} and \ref{fig2}]. The estimators are slightly more accurate under the assumption of an independent design where the differences are more substantial for the estimators of the second component. The differences between the uncorrelated and correlated case are even more visible
 for model \eqref{sig3}, for which the results are displayed in Figure \ref{fig3}. Here we observe that the first component is not estimated accurately
in a neighborhood of the origin. This is in accordance with our theoretical analysis, because the first component in model \eqref{sig3} does not satisfy the assumptions
made in Section \ref{sec3}. Consequently, the resulting estimates of the first component are biased in a neighbourhood of the origin. On the other hand, the second component  satisfies
 these assumptions and  the right
 panels of Figure \ref{fig3}  show  that the second component can be estimated with similar precision as in model \eqref{sig1} and \eqref{sig2}.

In order to compare the new method with the marginal integration method proposed in \cite{hilbisdet13} we  finally display  in Table \ref{tab1} the simulated integrated mean squared error
of both estimators for the models \eqref{sig1} - \eqref{sig3}. We observe  in the case of independent predictors  that  the backfitting approach yields an improvement of $50\%$
with respect to the integrated mean squared error criterion. Moreover, in the situation of dependent predictors as considered in \eqref{des2} the improvement is
even more substantial and varies between a factor $3$ and $4$.
 We expect that the advantages of the backfitting methodology are even larger with an increasing dimension of the predictor
$\textbf{X}$.

\begin{figure}[hbt!]
   \includegraphics[width=8cm]{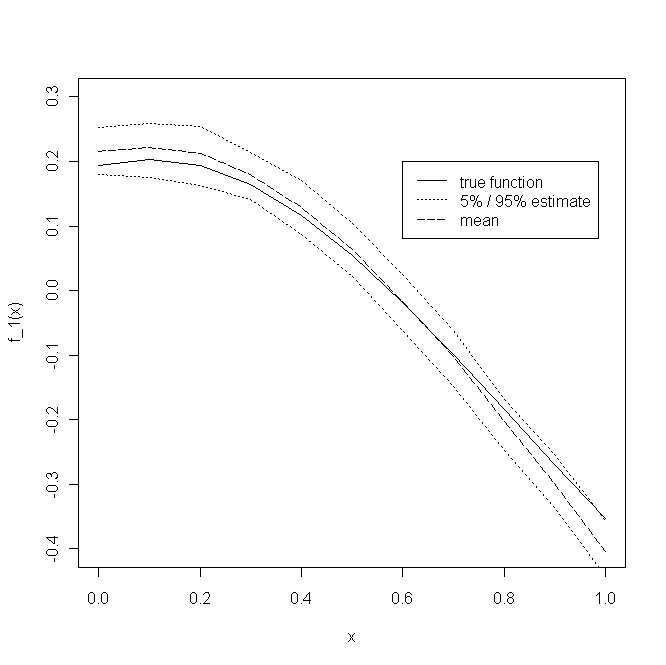}
 ~~~
    \includegraphics[width=8cm]{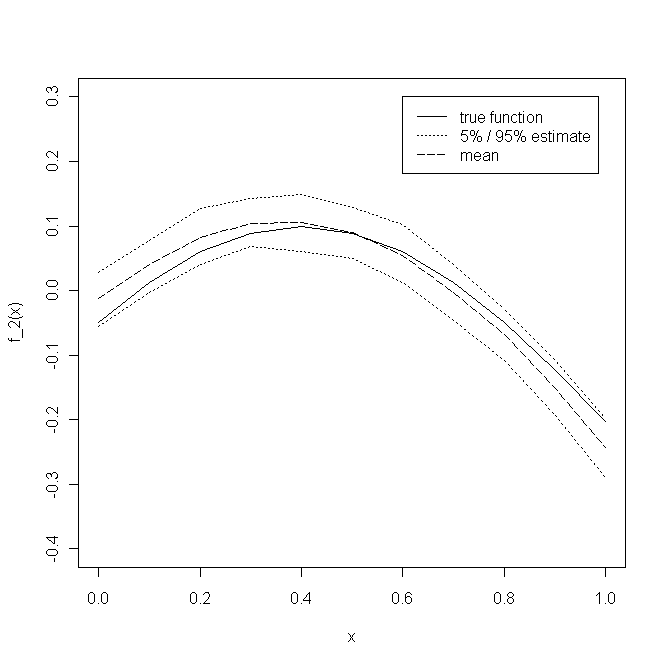} \\
 ~~~
        \includegraphics[width=8cm]{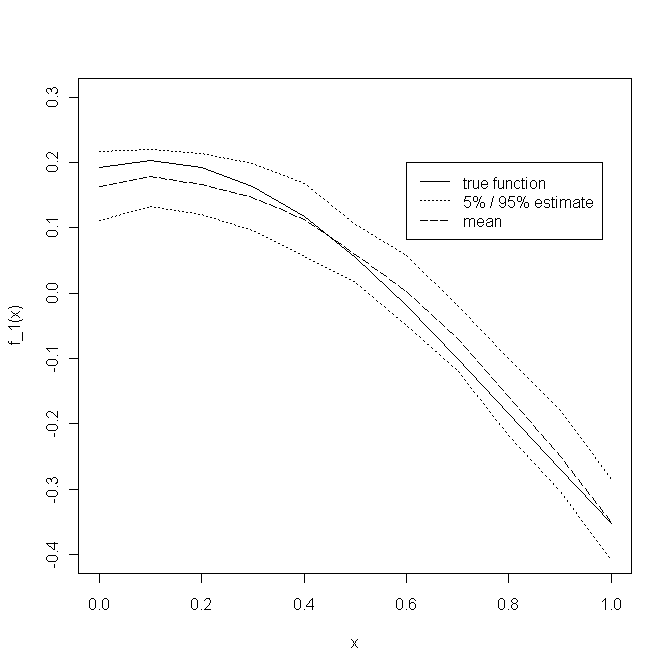}
 ~~~
    \includegraphics[width=8cm]{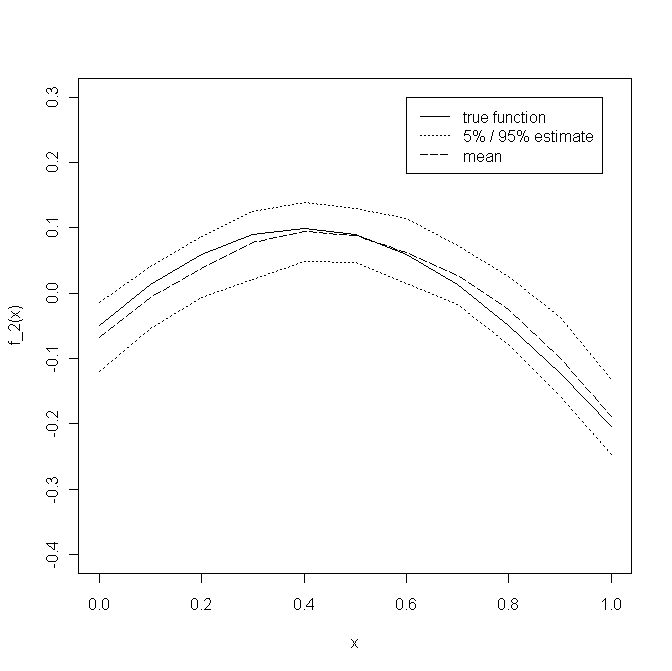}
   \caption{\label{fig1} \it Simulated mean, $5\%$- and $95\%$  quantile of the backfitting estimate on the basis of $500$ simulation runs, where model is  given by
  \eqref{sig1} and  the design is  given by \eqref{des1} (upper panel) and  \eqref{des2} (lower panel).  Left part $\theta_1$; right part: $\theta_2$. }
\end{figure}

\begin{figure}[hbt!]
    \includegraphics[width=8cm]{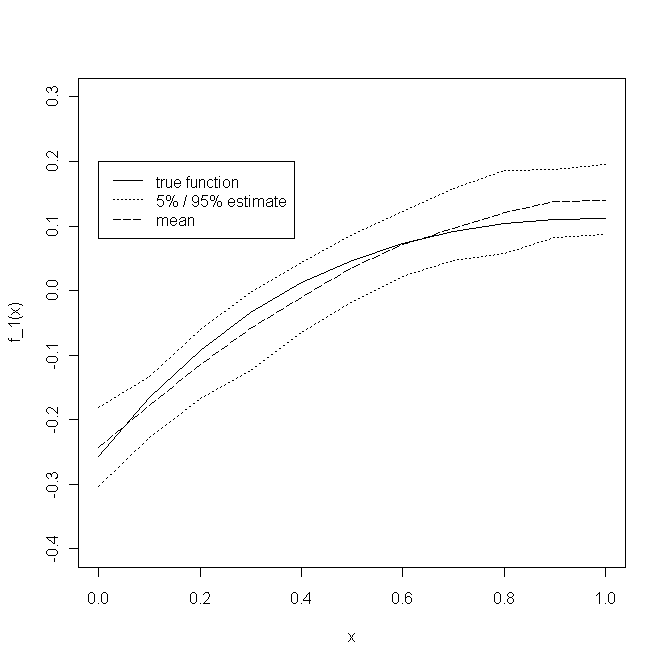}
 ~~~
     \includegraphics[width=8cm]{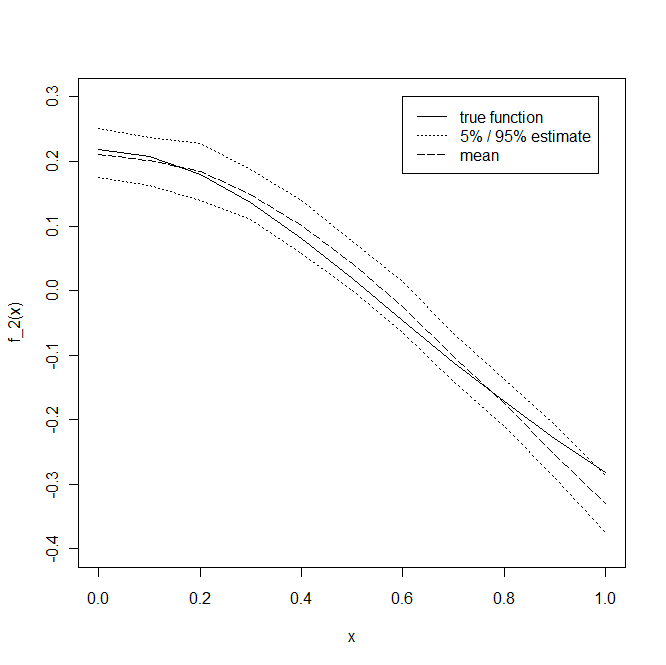} \\
 ~~~
         \includegraphics[width=8cm]{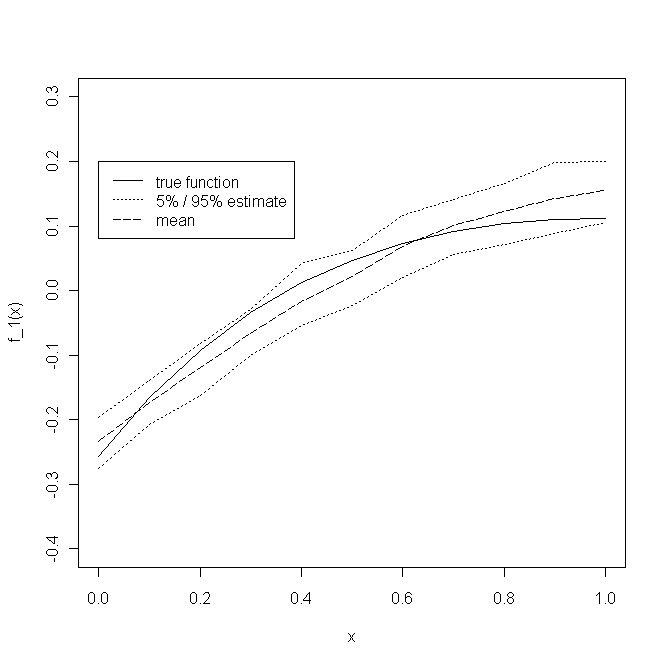}
 ~~~
   \includegraphics[width=8cm]{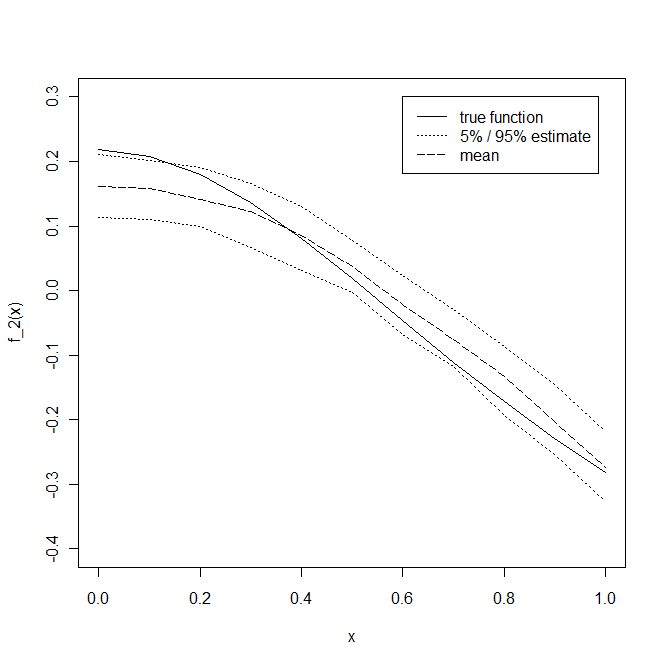}
   \caption{\label{fig2} \it Simulated mean, $5\%$- and $95\%$  quantile of the backfitting estimate on the basis of $500$ simulation runs, where model is  given by
  \eqref{sig2} and  the design is  given by \eqref{des1} (upper panel) and  \eqref{des2} (lower panel).  Left part $\theta_1$; right part: $\theta_2$. }
\end{figure}
\begin{figure}[hbt!]
    \includegraphics[width=8cm]{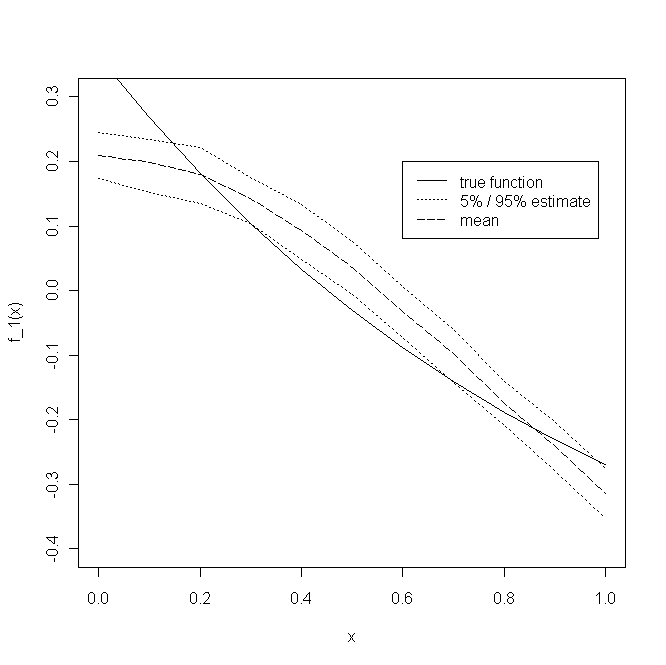}
 ~~~
    \includegraphics[width=8cm]{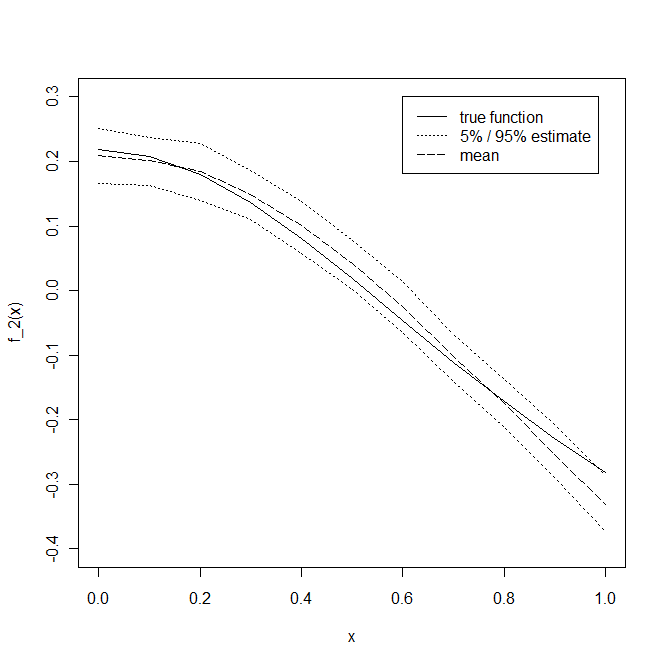} \\
 ~~~
           \includegraphics[width=8cm]{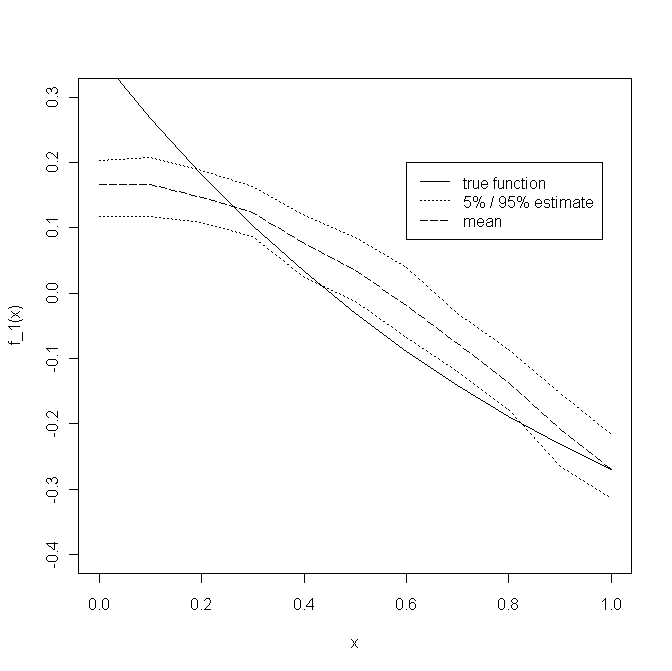}
 ~~~
     \includegraphics[width=8cm]{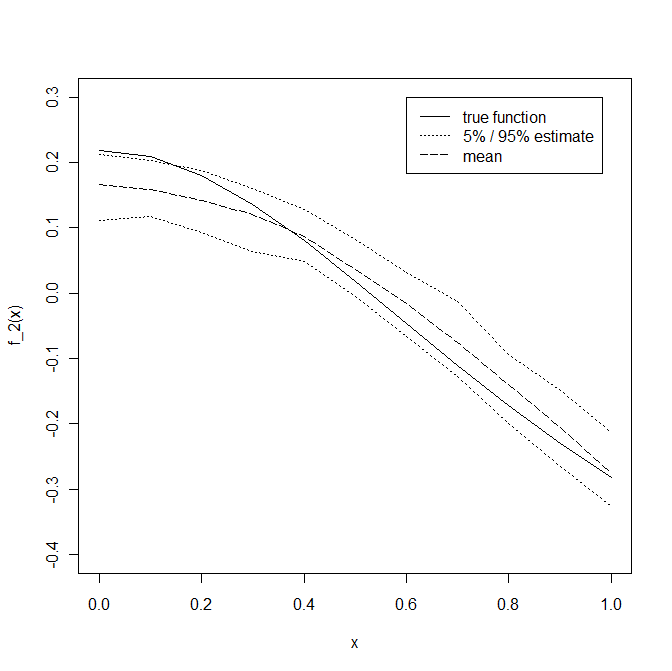}
   \caption{\label{fig3} \it Simulated mean, $5\%$- and $95\%$  quantile of the backfitting estimate on the basis of $500$ simulation runs, where model is  given by
  \eqref{sig3} and  the design is  given by \eqref{des1} (upper panel) and  \eqref{des2} (lower panel).  Left part $\theta_1$; right part: $\theta_2$. }
\end{figure}
\bigskip
\begin{table}[t]
\centering
{
\begin{tabular}{|| c|c|c||c|c||}
\hline
 design & \multicolumn{2}{c}  {\eqref{des1}} & \multicolumn{2}{||c||} {\eqref{des2}} \\ \hline
model & \eqref{sig1}  & \eqref{sig2}& \eqref{sig1}& \eqref{sig2}\\ \hline
$\hat \theta_1$ & 0.00179  & 0.00189& 0.00500& 0.00353\\
$\hat{\theta}_2$ & 0.00154  & 0.00258& 0.00488& 0.00345\\ \hline
$\hat{\theta}_1^{MI}$&0.00347  & 0.00365& 0.02219& 0.00934 \\
$\hat{\theta}_2^{MI}$ & 0.00311& 0.00354 &0.01917& 0.01092\\ \hline
\end{tabular}
}
\caption{{\it \label{tab1} Simulated mean integrated squared error of the smooth backfitting estimator $\hat{\theta}_j$  ($j=1,2$) proposed in this paper and of the
marginal estimator $\hat{\theta}_j^{MI}$ proposed by  \cite{hilbisdet13}.  }}
\end{table}

\vspace{.5cm}

{\bf Acknowledgements.}
The authors thank Martina
Stein and Alina Dette, who typed parts of this manuscript with considerable
technical expertise.
This work has been supported in part by the
Collaborative Research Center ``Statistical modeling of nonlinear
dynamic processes'' (SFB 823, Teilprojekt C1, C4) of the German Research Foundation
(DFG). \\

 \clearpage

\bibliography{bisdethil}

\begin{thebibliography}{}

\bibitem[Bertero et~al., 2009]{berbocdes2009}
Bertero, M., Boccacci, P., Desider{\`a}, G., and Vicidomini, G. (2009).
\newblock Image deblurring with {P}oisson data: {F}rom cells to galaxies.
\newblock {\em Inverse Problems}, 25(12):123006, 26.

\bibitem[Bissantz et~al., 2007]{bishohmunruy2007}
Bissantz, N., Hohage, T., Munk, A., and Ruymgaart, F. (2007).
\newblock Convergence rates of general regularization methods for statistical
  inverse problems.
\newblock {\em SIAM J. Num. Anal.}, 45:2610--2636.

\bibitem[Brillinger, 2001]{bril2001}
Brillinger, D.~R. (2001).
\newblock {\em Time Series Data Analysis and Theory}.
\newblock SIAM.

\bibitem[Carroll et~al., 2002]{carhaemam2002}
Carroll, R.~J., H{\"{a}}rdle, W., and Mammen, E. (2002).
\newblock Estimation in an additive model when the parameters are linked
  parametrically.
\newblock {\em Econometric Theory}, 18(4):886--912.

\bibitem[Cavalier, 2008]{cavalier2008}
Cavalier, L. (2008).
\newblock Nonparametric statistical inverse problems.
\newblock {\em Inverse Problems}, 24(3):034004, 19.

\bibitem[Diggle and Hall, 1993]{digglehall93}
Diggle, P.~J. and Hall, P. (1993).
\newblock A {F}ourier approach to nonparametric deconvolution of a density
  estimate.
\newblock {\em Journal of the Royal Statistical Society, Series B},
  55:523--531.

\bibitem[Engl et~al., 1996]{enghanneu1996}
Engl, H.~W., Hanke, M., and Neubauer, A. (1996).
\newblock {\em Regularization of inverse problems}, volume 375 of {\em
  Mathematics and its Applications}.
\newblock Kluwer Academic Publishers Group, Dordrecht.

\bibitem[Folland, 1984]{folland1984}
Folland, G.~B. (1984).
\newblock {\em Real Analysis - Modern Techniques and their Applications}.
\newblock Wiley, New York.

\bibitem[Gin{\'{e}} and Guillou, 2002]{gingui2002}
Gin{\'{e}}, E. and Guillou, A. (2002).
\newblock Rates of strong uniform consistency for multivariate kernel density
  estimators.
\newblock {\em Annales de l'Institut Henri Poincar{\'{e}} (B)
  Probabilit{\'{e}}s et Statistiques}, 38(6):907--921.

\bibitem[Hengartner and Sperlich, 2005]{henspe2005}
Hengartner, N.~W. and Sperlich, S. (2005).
\newblock Rate optimal estimation with the integration method in the presence
  of many covariates.
\newblock {\em Journal of Multivariate Analysis}, 95(2):246--272.

\bibitem[Hildebrandt, 2013]{hildebrandt2013}
Hildebrandt, T. (2013).
\newblock {\em Additive Modelle im inversen Regressionsproblem mit
  Faltungsoperator.}
\newblock PhD thesis, Fakult{\"a}t f{\"u}r Mathematik, Ruhr-Universit{\"a}t
  Bochum, Germany.

\bibitem[Hildebrandt et~al., 2013]{hilbisdet13}
Hildebrandt, T., Bissantz, N., and Dette, H. (2013).
\newblock Additive inverse regression models with convolution-type operators.
\newblock Submitted for publication,
  \url{http://www.ruhr-uni-bochum.de/mathematik3/research/index.html}.

\bibitem[Kaipio and Somersalo, 2010]{kaisom2005}
Kaipio, J. and Somersalo, E. (2010).
\newblock {\em Statistical and Computational Inverse Problems.}
\newblock Springer, Berlin.

\bibitem[Kammler, 2007]{kammler}
Kammler, D.~W. (2007).
\newblock {\em A first course in Fourier Analysis}.
\newblock Cambridge University Press.

\bibitem[Linton and Nielsen, 1995]{linnie1995}
Linton, O.~B. and Nielsen, J.~P. (1995).
\newblock A kernel method of estimating structured nonparametric regression
  based on marginal integration.
\newblock {\em Biometrika}, 82(1):93--100.

\bibitem[Mair and Ruymgaart, 1996]{mairuy1996}
Mair, B.~A. and Ruymgaart, F.~H. (1996).
\newblock Statistical inverse estimation in {H}ilbert scales.
\newblock {\em SIAM J. Appl. Math.}, 56:1424--1444.

\bibitem[Mammen et~al., 1999]{mamlinnie1999}
Mammen, E., Linton, O.~B., and Nielsen, J. (1999).
\newblock The existence and asymptotic properties of a backfitting projection
  algorithm under weak conditions.
\newblock {\em Annals of Statistics}, 27(5):1443--1490.

\bibitem[Nielsen and Sperlich, 2005]{niespe2005}
Nielsen, J.~P. and Sperlich, S. (2005).
\newblock Smooth backfitting in practice.
\newblock {\em Journal of the Royal Statistical Society, Ser.\ B},
  67(1):43--61.

\bibitem[Saitoh, 1997]{saitoh1997}
Saitoh, S. (1997).
\newblock {\em Integral Transforms, Reproducing Kernels and their
  Applications}.
\newblock Longman, Harlow.

\bibitem[van~der Vaart, 1998]{vaart1998}
van~der Vaart, A.~W. (1998).
\newblock {\em Asymptotic Statistics. Cambridge Series in Statistical and
  Probabilistic Mathematics}.
\newblock Cambridge, Cambridge University Press.

\end{thebibliography}

 \clearpage

\section{Appendix: Proof of Theorem \ref{theo1}} \label{sec5}

\def\theequation{5.\arabic{equation}}
\setcounter{equation}{0}

Let $p$ denote the density of the transformed predictor $(Z_{1,1},\dots,Z_{1,d})^T$.  It is shown in \cite{mamlinnie1999} that the smooth backfitting algorithm \eqref{backalg} produces a sequence
of estimates $(\hat g_1^{*(r)}, \ldots , \hat g_d^{*(r)})_{r=0,1,\ldots }$
converging in $L^2(p)$ with geometric rate  to a vector $(\overline{g}_1, \ldots  ,\overline{g}_d)$ which satisfies the system of equations
\begin{eqnarray}
\overline{g}_j (z_j) &=& \hat{g}^*_j(z_j)-\sum_{k\not =j} \int \overline g_k(z_k)\Big[ \frac{\hat{p}_{jk}(z_j,z_k)}{\hat{p}_j(z_j)}-\hat{p}_{k,[j+]}(z_k)\Big]dz_k-g^*_{0,j}
~~~~j=1,\ldots , d, \label{bfsol}
\end{eqnarray}
where $g^*_{0,j}$ is defined in  \eqref{g0j}.
Therefore the asymptotic properties of the smooth backfitting operator can be investigated replacing  in
\eqref{gjest} the random variables $U_{k,j}$ defined in \eqref{ukj} by their theoretical counterparts
\begin{eqnarray*}
\tilde U_{k,j} = Y_k -\sum_{\substack{i=1 \\ i\not=j}}^d \tilde{g}_i(X_{k,i}) - \hat g_0,
\end{eqnarray*}
         where $\tilde {g}_i(X_{k,i}) = \bar{g}_i(Z_{k,i})$
 $(i=1,\dots d; \ k=1,\dots,N)$ and $\tilde g_i = \bar {g}_i \circ F \ (i=1,\dots,d$). This yields the representation
\begin{eqnarray}
\tilde U_{k,j} = g_j(X_{k,j}) + \epsilon_k + \sum_{\substack{i=1 \\ i\not=j}}^d (g_i(X_{k,i})- \tilde  {g}_i(X_{k,i}))  =  g_j(X_{k,j}) + \epsilon_k + B_{j,k,N}, \label{bjkn}
\end{eqnarray}
where the last equality defines the random variables $B_{j,k,N}$ in an obvious manner. The results of \cite{mamlinnie1999} imply
\be \label{best}B_{j,k,N} = O_p(N^{-1/5})  \ee
uniformly with respect to $j \in \{1,\dots,d\}$ and $k \in \{1,\dots,N \}$.
\medskip

The assertion of Theorem \ref{theo1} is now proved in four steps establishing the following statements:
\begin{eqnarray} \label{lemEBFRD}
b_{\hat{\theta}_j}(x_j) &=&\mathbb{E}[\hat{\theta}_j(x_j)] - \theta_j(x_j)  = o(h^{s-1}) \\
\label{lemvarBFRD}
\mbox{Var}(\hat{\theta}_j(x_j)) &=& V _{N,j} (1+o(1)) \\
V_{n,j} && \quad \mbox{satisfies   (\ref{VNjung})}  \label{lemvertBFRD}\\
|\mbox{cum}_l(V_{N,j}^{-1/2}\hat{\theta}_j (x_j))|& =& o(1) ~~\mbox{for all }~l\ge 3
 \label{bound}
\end{eqnarray}
where $V _{N,j}$  is the normalizing factor   defined in \eqref{VNj} and cum$_l$ denotes the $l$th cumulant [see \cite {bril2001}].

\medskip
\textbf{\emph{Proof of (\ref{lemEBFRD}):}}  We first determine the expectation of the estimator  $\hat{\theta}_j$ observing
that the estimator $\hat \theta_j$ is linear, i.e.
\begin{eqnarray}
\hat{\theta}_j(x_j) &=&
\sum_{k=1}^N  w_{j,N} (x_j,X_{k,j}) \tilde U_{k,j}, \label{thetaaddrdbf}
\end{eqnarray}
where the weights $w_{j,N}(x_j,X_{k,j})$
are defined by
\begin{eqnarray}
w_{j,N}(x_j,X_{k,j}) = \frac{1}{ 2\pi Nh} \int_{\mathbb{R}} e^{-iw (x_j-X_{k,j})/h}  \frac{\Phi_K(w) }{\Phi_{\psi_{j}}(\frac{w}{h})} dw \frac {1} { \max \{ f_j (X_{k,j}), f_j(\frac {1}{a_n}) \} },
\label{wnBF}
\end{eqnarray}
and we have replaced the quantities $U_{k,j}$ by $\tilde U_{k,j}$ as described at the beginning of the proof.
This representation  gives
\begin{eqnarray} \label{decomp}
\mathbb{E}[\hat{\theta}_j(x_j)] &=& E_1+E_2,
\end{eqnarray}
where the terms ${E}_1$ and ${E}_2$ are defined by
\begin{eqnarray} \label{e1}
E_1 &=& \mathbb{E}\Bigl[ \sum_{k=1}^N g_j(X_{k,j}) w_{j,N}(x_j,X_{k,j})\Bigr],  \qquad E_2 =   \mathbb{E}\Bigl[\sum_{k=1}^N B_{j,k,N} w_{j,N}(x_j,X_{k,j})\Bigr].
\end{eqnarray}
Using the definition of  $B_{j,k,N}$ and  \eqref{best} the  term $E_2$ can be estimated as follows
\begin{eqnarray}
|E_2|
&\le& \mathbb{E}\Bigl[\sum_{k=1}^N  \sum_{\substack{i=1 \\ i\not=j}}^d |g_i(X_{k,i})-\tilde{g}_i(X_{k,i})| \max_k |w_{j,N} (x_j,X_{k,j})| \Bigr]  \label{E2}
\\ &\le& \frac{C}{h^{\beta_j+1}f_j(\frac{1}{a_N})}\mathbb{E}\Bigl[   \sum_{\substack{i=1 \\ i\not=j}}^d |g_i(X_{k,i})-\tilde{g}_i(X_{k,i})|  \Bigr] \le \frac{C}{N^{1/5}h^{\beta_j+1}f_j(\frac{1}{a_N})}=o(h^{s-1}), \nonumber
\end{eqnarray}
where we used the representation \eqref{wnBF} and Assumption (A4). The second inequality in \eqref{E2} follows from the fact that
\begin{eqnarray}
\mathbb{E}[|g_i(X_{k,i})-\tilde{g}_i(X_{k,i})|] = O(N^{-1/5}). \label{L1}
\end{eqnarray}
In order to establish this statement note that $g_i(X_{k,i})-\tilde{g}_i(X_{k,i})=O_P(N^{-1/5})$ (uniformly with respect to $k=1,...,N$). The proof of the $L^1$-convergence follows along the lines of the proof of the stochastic convergence in \cite{mamlinnie1999}. Here one additionally shows in each step of the backfitting iteration  stochastic convergence and $L^1$- convergence [see \cite{hildebrandt2013} for details].
\\ Similarly, we obtain from the definition of the weights  $w_{j,N} (x_{j},X_{k,j})$  in \eqref{wnBF} the representation
\begin{eqnarray} \label{E1}
E_1  &=&  \frac{1}{2\pi h} \int_{\mathbb{R}} g_j(y)   \int_{\mathbb{R}} e^{-iw (x_j-y)/h} \frac{\Phi_K(w)}{\Phi_{\psi_j}(\frac{w}{h})} dw \frac{f_j(y)}{\max\{ f_j(y),f_j(\frac{1}{a_N})\}} dy
\\ \nonumber
&=& \frac{1}{ 2\pi h} \int_{\mathbb{R}} \Phi_{g_j}\Bigl(\frac{w}{h}\Bigr)e^{-iw x_j/h} \frac{\Phi_K(w)}{\Phi_{\psi_j}(\frac{w}{h})} dw
\\ \nonumber
&& - \frac{1}{ 2\pi h}\int_{\mathbb{R}} g_j(y)   \int_{\mathbb{R}} e^{-iw (x_j-y)/h} \frac{\Phi_K(w)}{\Phi_{\psi_j}(\frac{w}{h})} dw \Bigl(1- \frac{f_j(y)}{\max\{ f_j(y),f_j(\frac{1}{a_N})\}} \Bigr) dy ~,
\\
\nonumber  &=& \theta_j(x_j) -F_1 -F_2,
\end{eqnarray}
where the terms $F_1$ and $F_2$ are defined by
\begin{eqnarray*}
F_1&=&  \frac{1}{ 2\pi h} \int_{\mathbb{R}} \Phi_{\theta_j}\left(\frac{w}{h}\right)e^{-iw x_j/h}\left(1- \Phi_K(w)\right) dw,
\\ F_2  &= & \frac{1}{ 2\pi h} \int_{\mathbb{R}} g_j(y)   \int_{\mathbb{R}} e^{-iw (x_j-y)/h} \frac{\Phi_K(w)}{\Phi_{\psi_j}(\frac{w}{h})} dw \Bigl(1- \frac{f_j(y)}{\max\{ f_j(y),f_j(\frac{1}{a_N})\}} \Bigr) dy,
\end{eqnarray*}
respectively. The term $F_1$ can be estimated using Assumption (A6), that is
\begin{eqnarray*}
|F_1 |
&\le& \frac{1}{ 2\pi h} \int_{\mathbb{R}} |\Phi_{\theta_j}\left(\frac{w}{h}\right) |1- \Phi_K(w)| dw  \le    \frac{1}{ \pi h} \int_{[-b,b]^c} |\Phi_{\theta_j}\left(\frac{w}{h}\right)| dw  \\
& \le  &  \frac{1}{\pi} \int_{[-b/h,b/h]^c} \frac{1}{| y |^{s-1}} | y |^{s-1}|\Phi_{\theta_j}(y)| dy
\\   &\le&  \frac{h^{s-1}}{b^{s-1}\pi} \int_{[-b/h,b/h]^c} | y |^{s-1}|\Phi_{\theta_j}(y)| dy
=   o(h^{s-1}),
\end{eqnarray*}
while the term $F_2$ is estimated  similarly, using Assumption (A4),   (A7) and (A8) that is
\begin{eqnarray*}
|F_2 |&\le&  \frac{1}{2\pi h} \int_{\mathbb{R}} |g_j(y)|   \int_{\mathbb{R}}  \frac{|\Phi_K(w)|}{|\Phi_{\psi_j}(\frac{w}{h})|} dw\Bigl|1- \frac{f_j(y)}{\max\{ f_j(y),f_j(\frac{1}{a_N})\}} \Bigr| dy
\\ &\le& \frac{1}{2\pi h} \int_{([-1/a_N,1/a_N])^c} |g_j(y)| dy  \int_{\mathbb{R}}  \frac{|\Phi_K(w)|}{|\Phi_{\psi_j}(\frac{w}{h})|} dw
~=~ O\left( \frac{a_N^r}{h^{1+\beta_j}} \right)
~=~   o(h^{s-1}).
\end{eqnarray*}
From these estimates and \eqref{E1} we obtain
$
E_1 = \theta_j(x_j) +o\left(h^{s-1}\right),
$
and the assertion \eqref{lemEBFRD} now follows from the decomposition \eqref{decomp} and \eqref{E2}.

\medskip

\textbf{\emph{Proof of  \eqref{lemvarBFRD}:} }Using standard results for cumulants [see \cite{bril2001}]  the variance of the estimate $\hat \theta_j$
can be calculated as
\begin{eqnarray}
\mbox{Var} (\hat{\theta}_j(x_j)) = S_1 +  S_2 +  S_3 +  2 S_4 +  2 S_5+  2 S_6,
\end{eqnarray}
where
\begin{eqnarray}
S_1  &=& \sum_{k=1}^N\sum_{l=1}^N \mbox{cum}\big( \epsilon_k w_{j,N}(x_j,X_{k,j}), \epsilon_l \overline{w_{j,N}(x_j,X_{l,j})}\big) \label{varber}  \nonumber \\
S_2  &=& \sum_{k=1}^N\sum_{l=1}^N \mbox{cum}\big( g_j(X_{k,j}) w_{j,N}(x_j,X_{k,j}), g_j(X_{l,j})\overline{w_{j,N}(x_j,X_{l,j})}\big) \nonumber \\
S_3  &=& \sum_{k=1}^N\sum_{l=1}^N \mbox{cum}\big( B_{j,k,N}w_{j,N}(x_j,X_{k,j}), B_{j,l,N}\overline{w_{j,N}(x_j,X_{l,j})}\big) \nonumber\\
S_4  &=& \sum_{k=1}^N\sum_{l=1}^N \mbox{cum}\big( \epsilon_{k}w_{j,N}(x_j,X_{k,j}), g_j(X_{l,j})\overline{w_{j,N}(x_j,X_{l,j})}\big) \nonumber \\
S_5  &=& \sum_{k=1}^N\sum_{l=1}^N \mbox{cum}\big( \epsilon_{k}w_{j,N}(x_j,X_{k,j}), B_{j,l,N}\overline{w_{j,N}(x_j,X_{l,j})}\big) \nonumber \\
S_6  &=&  \sum_{k=1}^N\sum_{l=1}^N \mbox{cum}\big( g_j(X_{k,j})w_{j,N}(x_j,X_{k,j}), B_{j,l,N}\overline{w_{j,N}(x_j,X_{l,j})}\big). \nonumber
\end{eqnarray}
It is easy to see that $S_4=0$   because of $\mathbb{E}[\varepsilon_k]=0$ and the independence of $\varepsilon_k$ and ${\bf X}_k$.
We will show that the first two terms $S_1$ and $S_2$ determine the variance and that the terms $S_3,S_5$ and $S_6$ are of smaller order. For a proof of the latter result we concentrate on the sixth term because the results for the terms $S_3$ and $S_5$ can be treated analogously.
\\{As $\epsilon_k$, $\epsilon_l$, $X_{k,j}$ and $X_{l,j}$ are independent for $k \not = l$ the term $S_1$ can be written as
\begin{eqnarray*}
N \mbox{cum}\big( \epsilon_k w_{j,N}(x_j,X_{k,j}), \epsilon_k \overline{w_{j,N}(x_j,X_{k,j})}\big)
&=&
N \mbox{cum}\big(\epsilon_k,\epsilon_k \big)\mbox{cum}\big(w_{j,N}(x_j,X_{k,j}),\overline{w_{j,N}(x_j,X_{k,j})}\big)
\\ &+&
N \mbox{cum}\big(\epsilon_k,\epsilon_k \big)\mbox{cum}\big(w_{j,N}(x_j,X_{k,j}) \big)\mbox{cum}\big( \overline{w_{j,N}(x_j,X_{k,j})}\big),
\end{eqnarray*}
where we used the product theorem for cumulants and $\mathbb{E}[\epsilon_k]=0$.
Now a straightforward calculation gives
\begin{eqnarray*}
S_1
=  \frac{\sigma^2}{Nh^2(2\pi)^{2}}\int_{\mathbb{R}}  \Bigl|  \int_{\mathbb{R}} e^{-iw(x_j- y)/h}
\frac{\Phi_K(w)}{\Phi_{\psi_j}(\frac{w}{h})} dw\Bigr|^2 \frac{f_j(y)}{\max\{ f_j(y),f_j(\frac{1}{a_N})\}^2} dy \cdot (1+o(1)).
\end{eqnarray*}
The second summand in \eqref{varber} can be calculated in the same way and we obtain
\begin{eqnarray*}
S_2 =\frac{1}{Nh^2(2\pi)^{2}}\int_{\mathbb{R}}  \Bigl|  \int_{\mathbb{R}} e^{-iw(x_j- y)/h}
\frac{\Phi_K(w)}{\Phi_{\psi_j}(\frac{w}{h})} dw\Bigr|^2 \frac{g^2_j(y)f_j(y)}{\max\{ f_j(y),f_j(\frac{1}{a_N})\}^2} dy \cdot (1+o(1)).
\end{eqnarray*}
In a last step we investigate the sixth summand of \eqref{varber} (the other terms $S_3$ and $S_5$ are treated in the same way). By the product theorem and the definition of the cumulants we obtain for this term
\begin{eqnarray*}
S_6 &=& - \sum_{k\not = l} \sum_{ \substack{ i=1 \\ i \not = j}}^d {\rm Cov}\big(g_j(X_{k,j})w_{j,N}(x_j,X_{k,j}), \overline{g}_i(F(X_{l,i})) w_{j,N} (x_j, X_{l,j} \big) \cdot (1+o(1)),
\end{eqnarray*}
where we used the definitions of $B_{j,l,N}=\sum_{i \not = j} (g_i(X_{l,i})-\tilde{g}_i(X_{l,i}))$ and $\overline{g}_i=\tilde{g}_i \circ F_i^{-1}$. We introduce the weights
\[q_{mj}(X_{l,i}) = \frac{L\big(\frac{F_j(X_{m,i})-F_j(X_{l,i})}{h_B}\big)}{\sum_{s=1}^N L\big(\frac{F_j(X_{s,i})-F_j(X_{l,i})}{h_B}\big)} \quad l,m=1,...,N; \quad i=1,...,d , \]
denote by
\begin{eqnarray}
 \hat{g}_i^*(F_i(X_{l,i})) &=& \sum_{m=1}^N q_{mi}(X_{l,i})Y_m \quad l=1,...,N;\quad i=1,...,d \label{NW}
\end{eqnarray}
the one-dimensional Nadaraya-Watson estimator   from the data $F_i(X_{1,i}),\dots, F_i(X_{N,i})$ evaluated at the point $F_i(X_{l,i})$ and define
\[ v_{mi}(X_{l,i},z_m) =\frac{\hat{p}_{im}(F_i(X_{l,i}),z_m)}{\hat{p}_i(F_i(X_{l,i}))}-\hat{p}_{m,[i+]}(z_m) \quad i,m=1,...,d; \quad l=1,...,N \]
as the integrand in equation \eqref{bfsol}. This yields for the term $S_6$ the decomposition
\begin{eqnarray} \label{s6}
{S_6}&=& (B-A )(1+o(1)),
\end{eqnarray}
where the terms $A$ and $B$ are defined by
\begin{eqnarray*} \label{aterm}
A&=&
\sum_{k \not =l} \sum_{ \substack{ i=1 \\ i \not = j}}^d  {\rm Cov} \big( g_j(X_{k,j})w_{j,N}(x_j,X_{k,j}), w_{j,N}(x_j,X_{l,j}) \sum_{m=1}^N q_{mi}(X_{l,i})Y_m  \big)  \\
\mbox{and} \\
B&=&
 \sum_{k \not=l} \sum_{ \substack{ i=1 \\ i \not = j}}^d \sum_{\substack{m=1 \\m \not = i}}^d {\rm Cov} \Bigl( g_j(X_{k,j})w_{j,N}(x_j,X_{k,j}), \bigl( \int \tilde{g}_m(z_m)v_{mi}(X_{l,i},z_m) dz_m+g_{0,i}^* \bigl)
w_{j,N}(x_j,X_{l,j}) \Bigr),
\end{eqnarray*}
respectively.
We start with the estimation of the term $A$ calculating each covariance separately, that is

\begin{eqnarray} \label{zerls6}
&& \Big|{ \rm Cov} \big( g_j(X_{k,j})w_{j,N}(x_j,X_{k,j}), w_{j,N}(x_j,X_{l,j}) \sum_{m=1}^N
q_{mi}(X_{l,i}) Y_m
\big)\Big|  \leq  \bigl( H_1 + H_2 \bigr) (1 + o(1)),
\end{eqnarray}
where the terms $H_1$ and $H_2$ are defined by
\begin{eqnarray*}
H_1 &=& \frac{1}{Nh_B}\Big|\sum_{r=1}^d\mathbb{E}\Big[  g_j(X_{k,j})w_{j,N}(x_j,X_{k,j}) L\Bigl(\frac{F_i(X_{k,i})-F_i(X_{l,i})}{h_B}\Bigr) g_r(X_{k,r}) \overline{w_{j,N}(x_j,X_{l,j})} \Big] \\
H_2 &=&  \frac{1}{Nh_B}\Big| \mathbb{E}\Big[g_j(X_{k,j})w_{j,N}(x_j,X_{k,j}) \Big] \mathbb{E} \Big[L\Bigl(\frac{F_i(X_{k,i})-F_i(X_{l,i})}{h_B}\Bigr) \sum_{r=1}^d g_r(X_{k,r}) \overline{w_{j,N}(x_j,X_{l,j})} \Big]
\end{eqnarray*}
and
we used the fact that the kernel density estimate
\[\frac{1}{Nh_B}\sum_{ m } L\Bigl(\frac{F_i(X_{m,i})-F_i(X_{l,i})}{h_B}\Bigr) = \hat{p}_i(F_i(X_{l,i})) \]
in the denominator of the Nadaraya-Watson estimate \eqref{NW} converges uniformly  to $1$ as $F_i(X_{l,i})$ is uniformly distributed on the interval $[0,1]$ [see \cite{gingui2002}].
We first investigate the term $H_1$ and obtain by a tedious calculation using assumption (A4) and (A9)
\begin{eqnarray*}
H_1 &\le& \frac {(1+o(1))} {N^3 h^2} \Big\vert \int\limits_{\mathbb{R}^2} \Big( \int\limits_{\mathbb{R}} g_j (t_j) \int\limits_{\mathbb{R}} e^{-iw (x_j - t_j)/h} \frac {\Phi_K (w)} {\Phi_{\psi_j} \Big( \frac w h \Big)} dw \frac {f_j(t_j) f_{irj} (t_i , t_r | t_j)} {\max \lbrace f_j(t_j) , f_j(1/a_N)\rbrace } dt_j \Big) \\
&\times& \int\limits_{\mathbb{R}} \Big( \int\limits_{\mathbb{R}} e^{-iw (x_j - s_j)/h} \frac {\Phi_K (w)} {\Phi(w/h)} dw \frac {f_j(s_j) f_{ij} (t_i | s_j)} {\max \lbrace f_j (t_j), f_j(1/a_N)\rbrace} ds_j \Big) g_j(t_r) dt_i dt_r \Big\vert \\
&\le& \frac C {N^3 h^2} \int\limits_{\mathbb{R}^2} \Big \vert \int\limits_{\mathbb{R}} e^{-iw (x_j - t_j)/h} \frac {\Phi_K (w)} {\Phi_{\psi_j} (w/h)} dw \frac {f_j(t_j)} {\max \lbrace f_j(t_j), f_j(1/a_N)\rbrace} dt_j \Big \vert \\
&\times& \Big\vert \int\limits_{\mathbb{R}} e^{-iw (x_j - s_j)/h} \frac {\Phi_K (w)} {\Phi_{\psi_j} (w/h)} dw \frac {f_j(s_j)} {\max \lbrace f_j(t_j), f_j(1/a_N)\rbrace} ds_j \Big \vert U_{irj} (t_i , t_r) \eta_{ij} (t_i) dt_i dt_r \\
&=& o \Big( \frac 1 {N^3h^{2\beta+1}} \Big)
\end{eqnarray*}
uniformly with respect to $k,l$.
A similar calculation yields
$$
H_2  \leq  \frac{1}{Nh_B}\Big| \frac{E_1}{N} \: \mathbb{E} \Big[L\Bigl(\frac{F_i(X_{k,i})-F_i(X_{l,i})}{h_B}\Bigr) \sum_{r=1}^d g_r(X_{k,r}) \overline{w_{j,N}(x_j,X_{l,j})} \Big] \Big| = o \Big( \frac 1 {N^3h^{2\beta+1}} \Big)
$$
(uniformly with respect to $k,l$) where we use the estimate \eqref{e1} in the first step. Consequently the term $A$ in \eqref{aterm} can be bounded by
$
A  =  o ( {1}/ {Nh^{2\beta+1}} )
$
 A tedious calculation using similar arguments yields for  the term $B = O ( {1} / {Nh^{2\beta+1}} )$ and by \eqref{s6} the sum $S_6$ is of the same order.  Moreover,   it will  be shown in  the proof of \eqref{lemvertBFRD} below that this
order is smaller than the order of the first two summands $S_1$ and $S_2$  in \eqref{varber} which gives
\[S_6 = O \Big( \frac{1}{Nh^{2\beta+1}} \Big) =o (S_j) \quad j=1,2.\]
A similar calculation for the terms $S_3$ and $S_5$ finally yields
\begin{eqnarray*}
\mbox{Var} (\hat{\theta}_j(x_j))&=& \frac{1}{Nh^2(2\pi)^{2}}\int_{\mathbb{R}}  \Bigl|  \int_{\mathbb{R}} e^{-iw(x_j- y)/h}
\frac{\Phi_K(w)}{\Phi_{\psi_j}(\frac{w}{h})} dw\Bigr|^2 \frac{(g_j^2(y)+\sigma^2)f_j(y)}{\max\{ f_j(y),f_j(\frac{1}{a_N})\}^2} dy  \times (1+o(1)) ~,
\\ &=& V_{N,j}  (1+o(1)) ~,
\end{eqnarray*}
which proves \eqref{lemvarBFRD}.

\bigskip

{\textbf{\emph{Proof of \eqref{lemvertBFRD}}}. As $g_j$ is bounded for all $j=1,...,d$ and $\max\{ f_j(y),f_j(\frac{1}{a_N})\}^2 \ge f_j(y)f_j(\frac{1}{a_N})$ the term $V_{N,j}$ defined in \eqref{VNj} can be estimated as follows
\begin{eqnarray*}
|V_{N,j}|&  \leq & \frac{C}{Nh(2\pi)^{2}f_j(\frac{1}{a_N})}\int_{\mathbb{R}}  \Bigl|  \int_{\mathbb{R}} e^{-iw(x_j/h- y)}
\frac{\Phi_K(w)}{\Phi_{\psi_j}(\frac{w}{h})} dw\Bigr|^2  dy  =
\frac{C}{Nh(2\pi)^{2}f_j(\frac{1}{a_N})}  \int_{\mathbb{R}}
\frac{|\Phi_K(w)|^2}{|\Phi_{\psi_j}(\frac{w}{h})|^2} dw ~,
\end{eqnarray*}
where $C$ is a constant and we used
Parseval's equality for the last identity [see \cite{kammler}].
Now assumption (A4) yields the upper bound, that is
$
|V_{N,j}| \leq {C}/{Nh^{1+2\beta_j}f_j(\frac{1}{a_N})}.
$
From the assumption  $f_j(x)^{-1} \ge C$ and again Parsevals equality we also get the lower bound
$
|V_{N,j}| \ge {C}/{Nh^{1+2\beta_j}},
$
which completes the proof of  the estimate  \eqref{VNjung}}.

\bigskip

\textbf{\emph{Proof of \eqref{bound}:}}  Observing the representation \eqref{thetaaddrdbf} the $l$th cumulant of the estimate $\hat{\theta}_j$ can be estimated as follows
\begin{eqnarray*}
|\mbox{cum}_l(\hat{\theta}_j(x_j))| &=&\Bigl|\sum_{k_1,...,k_l=1}^N \mbox{cum}\Bigl( \tilde{U}_{k_1,j}w_{j,N}(x_j,X_{k_1,j}),..., \tilde{U}_{k_l,j}w_{j,N}(x_j,X_{k_l,j})\Bigr)\Bigr| \leq G_1 + G_2,
\end{eqnarray*}
where the terms $G_1$ and $G_2$ are defined by
\begin{eqnarray*}
G_1 &=& \Bigl|\sum_{k_1,...,k_l=1}^N \mbox{cum}\Bigl( A_{k_1,j}w_{j,N}(x_j,X_{k_1,j}),..., A_{k_l,j}w_{j,N}(x_j,X_{k_l,j})\Bigr)\Bigr|
\\ G_2 &=& \Bigl|\sum_{k_1,...,k_l=1}^N \sum_{s=1}^l {l \choose s} \mbox{cum}\Bigl( B_{j,k_1,N}w_{j,N}(x_j,X_{k_1,j}), \ldots, B_{j,k_s,N} w_{j,N} (x_j, X_{k_s,j}), \\
&& A_{k_{s+1},j}w_{j,N}(x_j,X_{k_{s+1},j}),...,A_{k_l,j}  w_{j,N}(x_j,X_{k_l,j})\Bigr)\Bigr|
\end{eqnarray*}
and we introduce the notation $A_{k_i,j}=g_j(X_{k_i,j})+\epsilon_{k_i}$. Exemplarily we investigate the first term of this decomposition, the term $G_2$ is treated similarly. As the random variables $A_{k_1,j}w_{j,N}(x_j,X_{k_1,j})$ and $A_{k_2,j}w_{j,N}(x_j,X_{k_2,j})$ are independent for $k_1 \not = k_2$ and identically distributed for $k_1 =k_2$ it follows that
\begin{eqnarray*}
G_1 &=&  N\Bigl| \mbox{cum}_l  ( A_{k,j}w_{j,N}(x_j,X_{k,j})) \Bigr| \le  N\sum_{s=0}^l \dbinom{l}{s} \sum_{\substack{ \textbf{j} \in\{0,1\}^l \\ j_1+...+j_l=s}} \Bigl| \sum_\nu \prod_{k=1}^p \mbox{cum}(A_{ij}, ij\in \nu_k) \Bigr|,
\end{eqnarray*}
where we used the product theorem for cumulants  [see \cite{bril2001}] and
 the third sum extends over all indecomposable partitions of the table
\begin{table}[htb!]
\centering
\begin{tabular}{l c c}
$A_{i1}$ &  & $ A_{i2}$ \\
\vdots & & \vdots \\
$A_{i1}$\textcolor{white}{aaaa} &  & $ A_{i2}$ \\
& $A_{ij}$& \\
& \vdots & \\
& $A_{ij}$&
\end{tabular}
\end{table} \\
with
$A_{i1} =  \epsilon_1 $ ($ 1\le i \le s$), $ A_{i2} = w_{j,N}(x_j,X_{1,j}) $ ($ 1\le i \le s$)
and $ A_{ij} = g_j(X_{1,j})w_{j,N}(x_j,X_{1,j}) $ ($ s+1 \le i \le l$).
In order to illustrate how to estimate this expression we consider exemplarily the case $l=3$, where $G_1$ reduces to
\begin{eqnarray*}
G_1 &=& N\sum_{s=0}^3 \dbinom{3}{s} \sum_{\substack{ \textbf{j} \in\{0,1\}^3 \\ j_1+...+j_3=s}} \Bigl| \sum_\nu \prod_{k=1}^p \mbox{cum}(A_{ij}, ij\in \nu_k) \Bigr| .
\end{eqnarray*}
As $\varepsilon$ is independent of $X_1$ and has mean 0 the partitions in $G_1$ with $s=1$ vanish. The terms corresponding to $s=0,2,3$ contain only quantities of the form
\begin{eqnarray*}
&&\mbox{cum}_3(g_j(X_{1,j})w_{j,N}(x_j,X_{1,j})),
\\ &&\sigma^2 \mbox{cum}\big(w_{j,N}(x_j,X_{1,j}),w_{j,N}(x_j,X_{1,j}),g_j(X_{1,j})w_{j,N}(x_j,X_{1,j}) \big),
\\ && \sigma^2\mbox{cum}\big(w_{j,N}(x_j,X_{1,j})\big) cum\big(w_{j,N}(x_j,X_{1,j}),g_j(X_{1,j})w_{j,N}(x_j,X_{1,j}) \big),
\\ && \kappa_3 \mbox{cum}\big(w_{j,N}(x_j,X_{1,j}),w_{j,N}(x_j,X_{1,j}),w_{j,N}(x_j,X_{1,j})\big),
\\ && \kappa_3 \mbox{cum}\big(w_{j,N}(x_j,X_{1,j}),w_{j,N}(x_j,X_{1,j})\big)cum\big(w_{j,N}(x_j,X_{1,j})\big),
\\ && \kappa_3 \mbox{cum}\big(w_{j,N}(x_j,X_{1,j})\big) cum\big(w_{j,N}(x_j,X_{1,j})\big)cum\big(w_{j,N}(x^*,X_{1,j})\big),
\end{eqnarray*}
where $\kappa_3$ denotes  the third cumulant of $\epsilon_1$. As the inequality
\begin{eqnarray*}
\mathbb{E}\Big[|g_j(X_{1,j})w_{j,N}(x_j,X_{1,j}))|^{b_r} |w_{j,N}(x_j,X_{1,j}))|^{a_r - b_r} \Big] \le  \frac{C}{N^{a_r}h^{a_r(\beta_j+1)}f_j(\frac{1}{a_n})^{a_r}}
\end{eqnarray*}
holds for $0\le b_r\le a_r$ all terms can be bounded by
$
{C}/( {N^{3}h^{3(\beta_j+1)}f_j(\frac{1}{a_n})^{3}}).
$
This yields
\begin{eqnarray*}
N^{3/2}h^{3\beta_j+3/2}G_1 \le CN^{3/2+1}h^{3\beta_j+3/2}\frac{1}{N^{3}h^{3(\beta_j+1)}f_j(\frac{1}{a_n})^{3}}=o(1),
\end{eqnarray*}
where we used the conditions on the bandwidth in the last step.
Similar calculations for the general case show
\begin{eqnarray*}
N^{l/2}h^{l\beta_j+l/2}G_1 &=& O\big((N^{l/2-1}h^{l/2}f_j(\frac{1}{{a_N}})^l)^{-1}\big)=o(1)
\end{eqnarray*}
whenever $l \ge 3$. The term $G_2$ can be calculated in the same way, where for example one additionally has to use the estimate
$
\mbox{Cov}(B_{j,k,N},\epsilon_l) = O({1}/{N})
$
uniformly with respect to all $j=1,...,d$, and $k,l=1,...,N$, which follows from the definition of the backfitting estimator.

\end{document}